\documentstyle[aps,prl,multicol]{revtex}

\begin{document}
\input{epsf}
\draft
\preprint{}
\title{Shifting a Quantum Wire through a Disordered Crystal: 
Observation of Conductance Fluctuations in Real Space}
\author{ T. Heinzel, G. Salis, R. Held, S. L\"uscher, and K. Ensslin}
\address{Solid State Physics Laboratory, ETH Z\"{u}rich, 8093
Z\"{u}rich,  Switzerland\\}
\author{W. Wegscheider and M. Bichler}
\address{Walter Schottky Institut, TU M\"unchen, 85748 Garching, 
Germany, and\\
Institut f\"ur Angewandte und Experimentelle Physik, Universit\"at 
Regensburg, 93040 Regensburg, Germany\\}
\date{\today}
\maketitle
\begin{abstract}
A quantum wire is spatially displaced by suitable electric fields with respect to the scatterers 
inside a semiconductor crystal. As a function of the wire position, the low-temperature resistance 
shows reproducible fluctuations. Their characteristic temperature 
scale is a few hundred millikelvin, indicating a phase-coherent effect.  
Each fluctuation corresponds to a single scatterer entering or 
leaving the wire. This way, scattering centers can be counted one by 
one.
\end{abstract}
\pacs{PACS numbers: 72.20.Dp, 73,23.-b, 73.40.Kp}

\begin{multicols} {2}

\begin{figure}
\centerline{\epsfxsize=3.0in \epsfbox{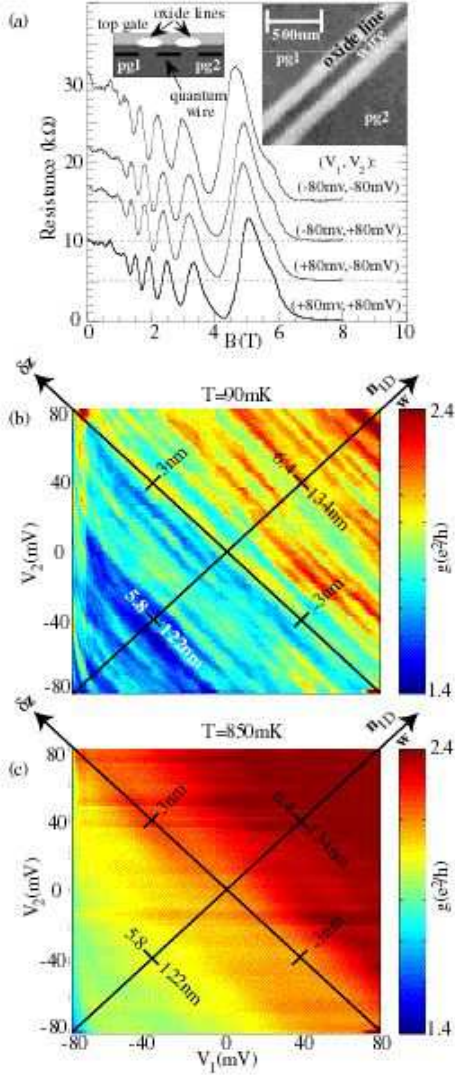}}
\caption{(a) Magnetoresistance for different planar 
gate voltages ($V_{1}$,$V_{2}$), corresponding to the corners in Figs. 
b and c. Curves are vertically offset by 
5\,k$\Omega$ each. Left inset: schematic cross section through the sample. The 2DEG is separated into wire 
and planar gates pg1 and pg2 by the oxide lines. The surface is
covered by a metallic top gate.  Right inset: atomic force 
microscope picture showing the surface 
topography before evaporation of the top gate. The oxidized 
regions are 8\,nm elevated above the heterostructure surface and 
appear as bright lines. (b):  Conductance $g$ at 
$T=90$\,mK and for $V_{tg}=0$, $B=0$ as a function of 
($V_{1}$,$V_{2}$). The data have been obtained by sweeping $V_{1}$ 
and changing $V_{2}$ in steps of 2 mV.
The two diagonal axes correspond to the electronic wire width $w$ (as 
well as to the electron density in the wire $n_{1D}$, measured in units of 
$10^{8}$m$^{-1}$), and to the wire 
displacement $\delta z$ (see text). (c):
The same measurement as in (b), taken at $T=850$\,mK.}
\label{CFFig1}
\end{figure}

\begin{figure}
\centerline{\epsfxsize=2.8 in \epsfbox{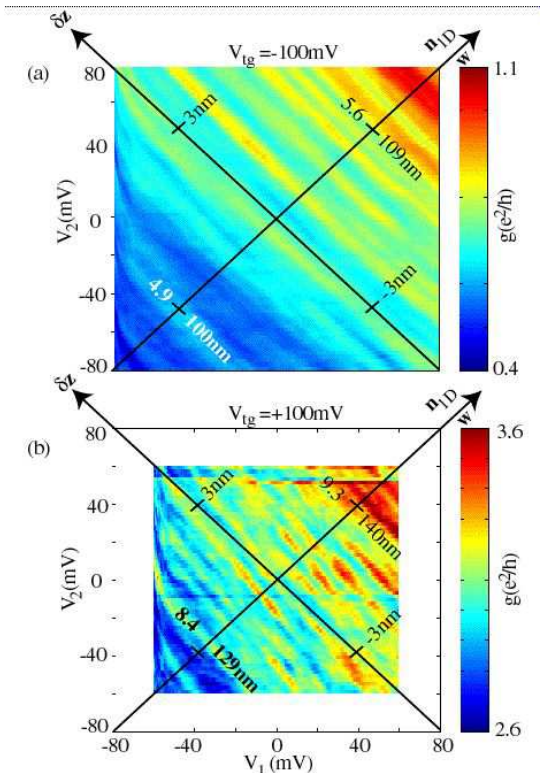}}
\caption{Wire conductance at $T=90$\,mK
in the ($V_{1}$, $V_{2}$)-plane for $V_{tg}$ = -100\,mV (a) and +100\,mV (b). 
The diagonal axes are defined as in Fig.~\ref{CFFig1}(b,c). At 
positive $V_{\rm tg}$, the fluctuation quasi-period in $\Delta z$ decreases 
significantly. At $V_{\rm tg}=100$\,mV, the break-down voltage of the 
planar gates decreases to $\pm $60\,mV, which reduces the scan range.}
\label{CFFig3}
\end{figure}

The mesoscopic regime in electronic transport is 
characterized by a sample size L comparable to the phase coherence length $\ell_{\phi}$ \cite 
{Zwerger98}. In this regime, electronic waves get reflected at the elastic 
scatterers and interfere with each other. As a consequence, the 
sample conductance $g$ depends on $\ell_{\phi}$ as well as on the configuration 
of the elastic scatterers, in contrast to the macroscopic regime, 
$\ell_{\phi}\ll$L. Theoretically, this effect can be treated by 
changing the scatterer configuration and studying the corresponding 
change in the conductance \cite {Feng86,Cahay88}. In the experiments, 
however, the conductance fluctuations (CF) are measured as a function of 
an externally controllable parameter, such as Fermi energy\cite{Licini85} 
or magnetic field\cite{Umbach84,Thornton87,Geim92}. 
The CF reflect the parametric modification of the interference pattern.
The experimental results are then connected to theory by using the ergodic 
hypothesis \cite{Altshuler91}, which states 
that averaging over the disorder in many samples is the same as 
changing a parameter that modifies the electronic properties 
at the Fermi level.  In completely phase coherent samples, the 
average CF amplitude is of the order of $e^{2}/h$ 
($e$ is the elementary charge and $h$ is Planck's constant), independent 
of material parameters. Therefore, they are also known as ``universal conductance fluctuations''
\cite{Lee87}, although true universality can be disturbed by various 
effects \cite{Onishi93,Nikolic94,Harris95,Zhu96}. The average amplitude 
is reduced not only when L gets larger than $\ell_{\phi}$, but also at non-zero 
temperature \cite{Beenakker91}.\\
Here, we report the experimental observation of CF 
as a function of the scatterer distribution inside a quasi-one-dimensional wire,
which can be shifted through the host crystal by gate 
electrodes.  Pronounced fluctuations in $g$ as a function of the displacement 
$\delta z$ of the wire perpendicular to the transport direction are 
found, which is the central result of 
this letter. We show that the quasi-period of the CF essentially reflects 
the displacement needed to shift the wire into or out of a single 
scatterer. \\
The sample is a Ga[Al]As heterostructure with the 
two-dimensional electron gas (2DEG) residing 37\,nm below the surface. 
 A 10\,$\mu$m wide Hall bar was defined by 
 optical lithography and wet chemical etching. Two oxide lines, 
 written with an atomic force microscope in the heterostructure 
 surface~\cite{Held98}, define 
 the wire of length $L=40$\,$\mu$m and lithographic width 
 of 150\,nm (inset in Fig.~\ref{CFFig1}(a)). In a final 
 fabrication step, the Hall bar was covered by a homogeneous Ti/Au top gate 
 electrode~\cite{Held99}. The sample was mounted in the mixing chamber of a 
 $^{3}$He/$^{4}$He-dilution refrigerator with a base temperature of 
 30\,mK. The sample chamber is carefully shielded from high frequency 
 noise. We estimate the electronic temperature to be about 
 $T = 90$\,mK. Under these 
 conditions  and with the top gate grounded, the 2DEG has a sheet density 
 of 4.3$\cdot 10^{15}$ m$^{-2}$ and a mobility of 
 102\,m$^{2}$/Vs. The Drude and quantum
 scattering lengths are $\ell_{D}=11.0$\,$\mu$m, 
 and $\ell_{q} = 0.46$\,$\mu$m, respectively, while the thermal length 
 is  $\ell_{T}=11.5$\,$\mu$m. \\
 A DC current of $2$\,nA was passed through 
 the quantum wire, and the voltage drop 
 was measured using  a four-terminal setup. Magnetic fields up to 
 $B$\,=\,$8$\,T were applied 
 perpendicular to the plane of the sample surface. From the autocorrelation function 
 of the CF as a function of $B$~\cite{Beenakker91}, we obtain 
 $\ell_{\phi} \approx$  7 $\mu$m. The wire can be tuned by voltages 
 $V_{\rm i}$, applied to two planar gates pgi (i=1,2), as well 
 as with a top gate voltage $V_{tg}$ (Fig.~\ref{CFFig1}(a)). Assuming a parabolic confinement, 
 we use the model by 
  Berggren et al.~\cite{Berggren87} to fit the positions of the Shubnikov-de 
  Haas minima, from which we obtain the one-dimensional electron density 
  $n_{1D}$ and the electronic wire width $w$. In Fig.~\ref{CFFig1}(a), the magnetoresistance of the 
 wire is shown for different planar gate voltages ($V_{1}$, $V_{2}$) with the top gate
 grounded. By changing 
 $V_{1}$ and/or $V_{2}$, $n_{1D}$  and $w$ can be tuned from
 ($n_{1D}=5.5\cdot10^8$\,m$^{-1}$, $w=116$\,nm) for $V_{1}$\,=\,$
 V_{2}$\,=\,$-80$mV to ($n_{1D}$\,=\,$6.7\cdot$10$^8$\,m$^{-1}$, $w$\,=\,$141$\,nm) for $V_{1}$\,=\,$
 V_{2}$\,=\,$+80$mV. Our scan range is 
 limited to
$\mid$$V_{i}$$\mid \leq$\,$80$\,mV, since a leakage current 
 across the oxide lines is detected for higher $\mid$$V_{i}$$\mid$. Within this range, the wire contains 8 spin-degenerate 
 one-dimensional subbands. Furthermore,  $R(B)$ and thus the fit parameters for ($V_{1},V_{2})$\,=\,$(-80$\,mV$, +80$\,mV) are identical to those for
 ($V_{1},V_{2})$\,=\,$(+80$\,mV$, -80$\,mV). Hence, by applying antisymmetric voltage changes 
to $V_{1}$ and $V_{2}$ (i.e. $\delta V_{1}$\,=\,$-\delta V_{2}$), the wire can effectively be shifted 
 through the semiconductor host while keeping $w$ and $n_{1D}$ 
 constant. A similar technique has been used already in order to 
 investigate the spatial location of noise-generating carrier traps in 
 quasi-ballistic channels\cite{Sakamoto95}, and to study the average 
 distribution of elastic scatterers of parabolic quantum wells in growth 
 direction\cite{Salis99}. The shift $\delta z$ can be estimated by 
 assuming that a voltage difference $\delta V_{i}$ (i=1,2) on one planar gate 
 moves the wire edge next to it by half the 
 change in wire width achieved when both gate voltages are changed 
 symmetrically by $\delta V_{1}$\,=\,$\delta V_{2}$. We find a lever 
 arm of $\delta z$/$\mid{\delta V_{i}\mid }$$\approx $$70$\,nm/V.
 Fig.~\ref{CFFig1}(b) shows the wire conductance in the plane spanned 
 by the planar gate voltages. The recording time for such a measurement 
 is 30 hours. Background charge 
 rearrangements, which appear as sudden, small shifts in the conductance pattern, are observed 
 on a time scale of $\approx$1 day. 
The diagonal axis running 
 from $(V_{1},V_{2})$=$(-80$\,mV$,-80$\,mV) to ($80$\,mV$,80$\,mV) corresponds to $\delta z$\,=\,$0$, 
 while the axis from 
  $(V_{1},V_{2})=(-80$\,mV$, +80$\,mV) to $(V_{1},V_{2})=(+80$\,mV$, -80$\,mV) corresponds to 
 constant $n_{1D}$ and $w$. Both axes have been calibrated by 
 measuring and fitting $R(B)$ for the points ($V_{1},V_{2})=(-$80$\,mV+m\cdot $20$\,mV$, 
 -80$\,mV$+n$\cdot$ 20$\,mV$) (m, n=0,$\ldots$,8). \\
 Reproducible fluctuations in the conductance are observed as a 
 function of both 
 $n_{1D}$ and $\delta z$. They are almost completely smeared out at $T\approx 850$\,mK 
 [Fig.~\ref{CFFig1}(c)], indicating a phase coherent origin. The overall conductance increases as 
 T is increased,  since the weak localization effect\cite{Beenakker91} is reduced. The characteristic 
 fluctuation period is $\approx  5 \cdot 
 10^{6}m^{-1}$ in $n_{1D}$  and $\approx  4 nm$ in $\delta z$. The characteristic 
 CF fluctuation amplitude 
 in both directions is $\delta g \approx$ 0.25 $e^{2}/h$, in 
 good agreement with the value expected from theory for $w\ll 
 \ell_{\phi}\approx \ell_{T} < L$ \cite{Beenakker87}, 
 i.e. $\delta 
 g\approx  \sqrt{12}(\ell_{\phi}/L)^{1.5}\cdot[1+(9/2\pi)(\ell_{\phi}/\ell_{T})^{2}]^{-0.5}e^{2}/h$, which gives  
 $\delta g\approx 0.22e^{2}/h$ for our parameters.\\ 
 In the following, we discuss the origin of CF as a function 
  of the displacement of the wire. \\
 Cahay et al.\cite{Cahay88} have studied theoretically CF as a single scatterer is 
 moved inside a quantum wire. In this case, the CF quasi-period 
was found to be of the order of the Fermi wave length $\lambda_{\rm F}$ of the lowest 
one-dimensional subband. In our sample,  $\lambda_{\rm F}\approx 30$\,nm, which 
is about one order of magnitude larger than 
the  observed quasi-period. Possibly, this theoretical model could be 
realized experimentally by using scanning probe microscopes to induce 
a scattering center in a quantum wire\cite{Eriksson96}. In our 
experiment, however, we do not 
scan individual scatterers with respect to the others inside the wire, 
but rather scan the 
conductive region through a rigid configuration of scatterers. Occasionally,  
a scatterer enters or exits the electronic wire. 
This changes the interference pattern of the electronic waves and 
influences the conductance. On average, the number of scatterers 
 inside the wire changes by one if the wire is 
displaced by $\Delta z=d^{2}/2L$ ($d$ is the relevant scattering 
length). The factor of two takes into account that 
scatterers may enter the wire on one side as well as exit it on the other. 
We find $\Delta z= 2.7$\,nm for $d=\ell_{q}$ and $\Delta z= 1.5$\,$\mu$m for $d=\ell_{D}$. This suggests that the observed quasi-period 
in $\delta z$  is caused by individual 
small-angle scatterers that the wire passes on its way through the 
crystal. As we will argue below, the actual situation is more complicated, 
since the two-dimensional quantum scattering length is not a good 
quantity to characterize the scattering inside the quantum wire. 
Rather, additional scattering is produced by the wire boundaries. 
This interpretation of the data in Fig.~\ref{CFFig1} is supported not 
only by a simple argument based upon a numerical 
simulation, but also by further, density-dependent measurements.\\ 
\begin{figure}
\centerline{\epsfxsize=2.5 in \epsfbox{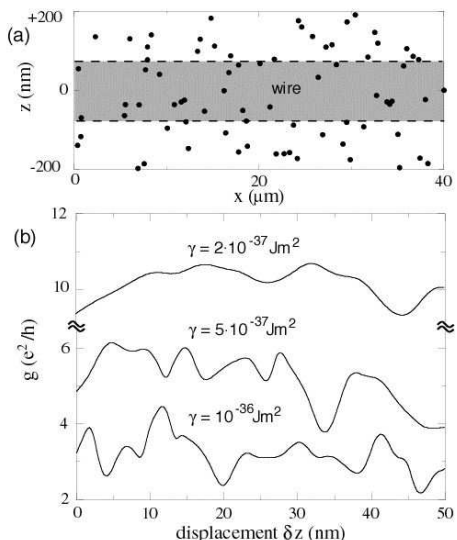}}
\caption{(a) Distribution of delta-shaped scatterers around the wire. The shaded region shows the wire position. 
(b) calculated conductance fluctuations g($\delta z$). With increasing 
scattering strength $\gamma$, g decreases and the quasi-period drops from 
15\,nm to about 5\,nm.} 
\label{CFFig2}
\end{figure}
We have modeled the wire conductance using the scattering-matrix 
approach of Ref.~\cite{Cahay88}. All scatterers are represented by identical delta-function potentials 
$\gamma\cdot\delta(x,z)$, located at random positions within an area 
of 400\,nm $\cdot$ 40\,$\mu$m. We choose their average distance equal to 
$\ell_{q}$ [Fig.~\ref{CFFig2}(a)].  The wire parameters (i.e. 
$w$, $\omega_{0}$, and $n_{1D}$) are taken from 
the data in Fig.~\ref{CFFig1}. Transport through the wire is assumed 
to be completely phase coherent. The conductance is obtained from the transmission part of the overall scattering 
matrix, which is composed of coherently combined
propagation matrices and single-impurity scattering matrices. 
The conductance as a function of the position of the wire, 
g($\delta z$), is obtained by shifting the conducting region through 
the rigid configuration of scatterers. Fig.~\ref{CFFig2}(b)
shows g($\delta z$) of a wire with 8 occupied modes ($w$\,=\,$130$\,nm, 
$\omega_{0}$\,=\,$3.1\cdot10^{12}$\,s$^{-1}$, $n_{\rm 
sc}$\,=\,$4.7\cdot10^{12}$\,m$^{-2}$) for 3  different values of 
$\gamma$. The calculated CF quasi-period
is of the same order as $\lambda_{\rm F}\approx 30$\,nm for small $\gamma$. 
By increasing $\gamma$ , however, it  can be reduced to the 
quasi-period observed experimentally. Note that in this simulation, the amplitude is of order 
$e^{2}/h$, since transport through the wire is completely phase coherent in our model.\\
We have also simulated g($\delta z$) when the number of scatterers 
inside the wire is 
kept fixed (not 
shown). Here, the quasi-period is of the order $\lambda_{\rm F}$, 
independent of $\gamma$. Furthermore, the quasi-period does not depend 
on the wire width at constant scatterer density. 
Our simulations thus indicate that only those 
scatterers that enter or exit the wire determine the observed 
quasi-period for large $\gamma$. As $\gamma$ is reduced, the 
quasi-period approaches $\lambda_{\rm F}$. We are not aware of any 
theoretical consideration on how the 
observed quasi-period should depend upon $\ell_{\phi}$. Qualitatively, 
however, we expect it to increase as $\ell_{\phi}$ gets smaller, as 
this is the case for the CF in magnetic fields and in electron density 
\cite{Beenakker91}. 
Consequently, we can consider the value for $\gamma$ that simulates 
the experiment as a lower limit for the scattering strength.  The experiment 
is simulated best for a value of $\gamma 
\approx 10^{-36}$Jm$^{2}$.  Assuming a scattering cross section with a 
radius $r_{sc}$ of the screening length for a 2DEG in GaAs \cite{Ando82} , i.e. 
$r_{sc}\approx a_{B}^{*}/2=5$nm¥ ($a_{B}^{*}$ denotes the effective Bohr 
radius in GaAs), we find as a lower limit a characteristic 
height of the scatterers, given by  $\gamma / (\pi r_{sc}^{2}) \approx 80$ meV. 
This corresponds to a strong scatterer. We 
conclude that in the wire region, scattering is enhanced. Presumably, 
the additional scattering comes from the roughness of the wire 
boundaries \cite{Nikolic94,Harris95,Zhu96,Thornton89,Ando92}. 
In addition, in the depleted regions underneath the oxide 
lines, screening is strongly reduced. \\
In order to test this picture, we have changed $w$ by applying top 
gate voltages [Fig.~\ref{CFFig3}]. \\
As the wire is narrowed to $w\approx 100nm$  ($V_{\rm tg}=-100$\,mV, 
Fig.~\ref{CFFig3}a), only 6 one-dimensional modes are occupied. The feature sizes 
 in $\delta z$ increase almost above our scan range. The fluctuation amplitude is 
 reduced. Note that the smearing of the CF as a function of $n_{1D}$, which 
 is determined by the reduction of $\ell_{phi}$, is weaker. Compared to the measurements in 
 Fig.~\ref{CFFig1}, the edges of the wire are $15$ nm further away from 
 the oxide lines. Consequently, the boundary roughness is smoothed and the quasi-period 
 increases, as seen qualitatively in Fig.~\ref{CFFig2}b). For positive 
 top gate voltages ($V_{\rm tg}=+100$\,mV, Fig.~\ref{CFFig3}b), $w$ increases only slightly, while $n_{1D}$ 
 increases rapidly. Here, 12 one-dimensional modes are occupied. The quasi period is decreased to about $2$nm, 
 which reflects the fact that here the wire is pushed very hard into the 
 oxidized region. Note that the observed dependence of the quasi-period 
 on the top gate voltage  is opposite to what is expected from 
 density dependent screening, since increasing $n_{1D}$ improves the 
 screening, which should result in larger quasi-periods. However, our 
 interpretation does not take into account the density-dependence of 
 $\ell_{\phi}$, which may play a role as well. 

In conclusion, we have reported the measurement of conductance 
fluctuations in a quantum 
wire as a function of its position with respect to a rigid scatterer 
configuration.  The 
fluctuation period decreases with increasing electron density. 
We have interpreted the period in real space in terms of individual scatterers 
entering or leaving the wire region, while its dependence on the 
width of the wire can be understood in terms of increased scattering 
at the wire boundaries. It will be particularly interesting to apply 
out technique to quasi-ballistic quantum point contacts as well as to 
single-mode quantum wires. We hope that our experiment will stimulate 
theoretical studies as well as new experiments, in  order to get a better understanding 
of conductance fluctuations in real space.\\
 We have enjoyed fruitful discussions with S.E. Ulloa, D. Loss and T. Ihn. This work was 
 supported by the Swiss National Science Foundation.\\

\end{multicols}
\end{document}